%% file: main_kky_v6_jcy2.tex
\documentclass[10pt,twocolumn,letterpaper]{article}

\usepackage[pagenumbers]{cvpr} 

\usepackage{graphicx}
\usepackage{amsmath}
\usepackage{amsthm}
\usepackage{amssymb}
\usepackage{booktabs}
\usepackage{mathtools}
\usepackage[linesnumbered,ruled,vlined]{algorithm2e}
\usepackage{latexsym}
\usepackage{graphicx}
\usepackage{multirow}
\usepackage{adjustbox}
\usepackage{makecell}
\usepackage{caption}
\usepackage{xcolor}
\definecolor{textblue}{rgb}{.2,.2,.7}
\definecolor{textred}{rgb}{0.54,0,0}
\definecolor{textgreen}{rgb}{0,0.43,0}
\usepackage{pifont}

\newcommand{\cmark}{\text{\ding{51}}}%
\newcommand{\xmark}{\text{\ding{55}}}%
\usepackage{listings}
\usepackage[pagebackref,breaklinks,colorlinks]{hyperref}

\usepackage[capitalize]{cleveref}
\crefname{section}{Sec.}{Secs.}
\Crefname{section}{Section}{Sections}
\Crefname{table}{Table}{Tables}
\crefname{table}{Tab.}{Tabs.}
\newtheorem{prop}{Proposition}

\def\cvprPaperID{11676} 
\def\confName{CVPR}
\def\confYear{2022}
\input{macros}

\title{Noise Distribution Adaptive Self-Supervised Image Denoising using \\ Tweedie Distribution and Score Matching}

\author{%
Kwanyoung Kim$^{1}$ \; Taesung Kwon$^{1}$ \; Jong Chul Ye$^{1,2,3}$\\
  \normalfont  $^1$ Department of Bio and Brain Engineering\\
  $^2$Kim Jaechul Graduate School of AI\\
    $^3$Deptartment of Mathematical Sciences\\
   Korea Advanced Institute of Science and Technology (KAIST) \\
\texttt{\{cubeyoung, star.kwon, jong.ye\}@kaist.ac.kr} \\
}

\begin{document}

\twocolumn[{%
	\renewcommand\twocolumn[1][]{#1}%
	\maketitle
	\begin{center}
		\centering
		\captionsetup{type=figure}
		\includegraphics[width=0.97\linewidth]{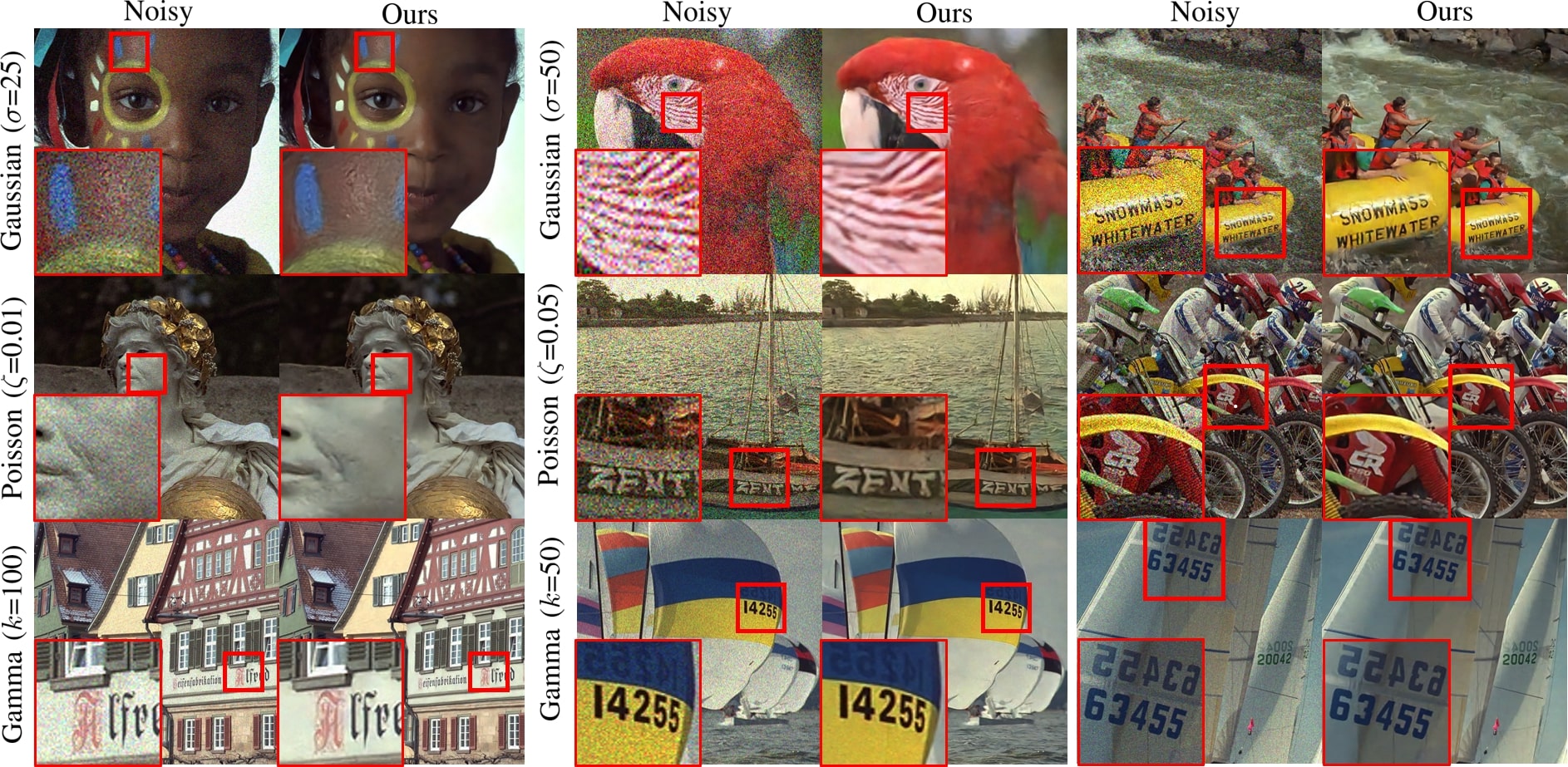}
		\captionof{figure}{Our image denoising results in Kodak dataset. Top: Gaussian noise ($\sigma$ =25, $\sigma =50$), Middle: Poisson noise ($\zeta$ = 0.01, 0.05), Bottom: Gamma noise ($k$ = 100, 50).}
	\end{center}%
}]
\begin{abstract}
Tweedie distributions are a special case of exponential dispersion models, which are often used  in classical
statistics as distributions for generalized linear models.
Here, we reveal that Tweedie distributions also play key roles in modern deep learning era, leading to a distribution independent
self-supervised image denoising  formula without clean reference images.
Specifically, by combining with the recent Noise2Score self-supervised image denoising approach and the saddle point approximation of
Tweedie distribution,
we can provide a general closed-form denoising formula that can be used for large classes of noise distributions without ever knowing the underlying
noise distribution.
Similar to the original Noise2Score, the new approach is composed of two successive steps: score matching
using perturbed noisy images, followed by a closed form image denoising formula via distribution-independent Tweedie's formula.
This also suggests a systematic algorithm to estimate the noise model and noise parameters for a given noisy image data set.
Through extensive experiments, we demonstrate that the proposed method can accurately estimate noise models and parameters, and provide the state-of-the-art self-supervised image denoising performance in the benchmark dataset and real-world dataset.

%
\vspace{-1em}
\end{abstract}

\begin{figure*}[!t]
	\centering
	\includegraphics[width=0.95\linewidth]{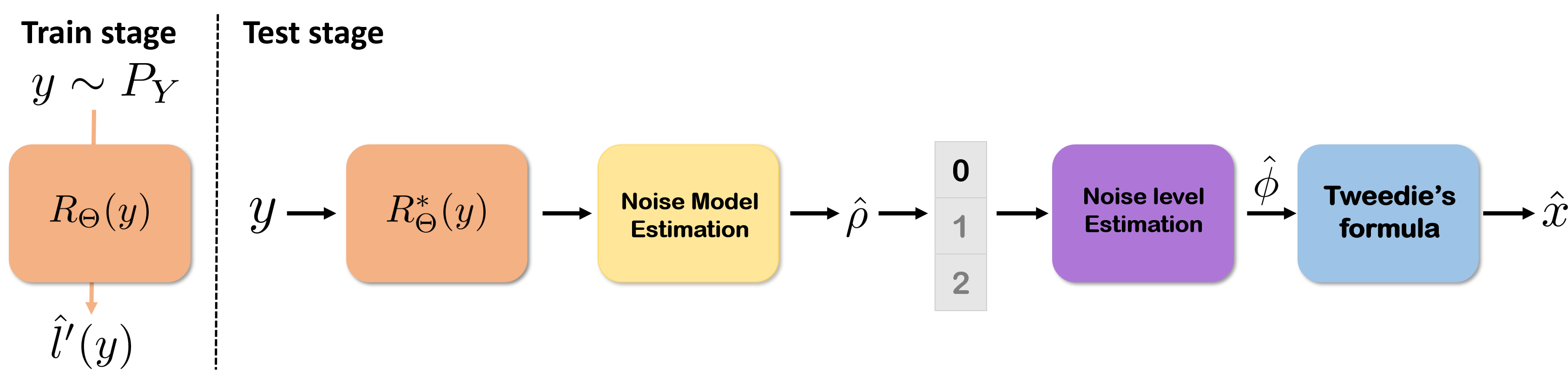}
	\caption{ Overall reconstruction flow of the proposed method, where the first step is the estimation of the score function $\hat{l}'$ by training a neural network $R_\Theta$, which is followed by the estimation of noise model and noise level to obtain final denoised results. During the training procedure, input images $y$ that belong to the distribution of train set $P_{Y}$, are sampled.}
	\label{fig:concept}
\end{figure*}

\section{Introduction}

Image denoising is a fundamental problem in low-level vision problems. 
Nowadays, typical supervised learning approaches easily outperform classical denoising algorithms such as Block-Matching and 3D filtering (BM3D)~\cite{dabov2006image} and Weighted Nuclear Norm Minimization (WNNM)~\cite{gu2014weighted}.  
Nonetheless, the supervised approaches are not practical  in many real-world applications as they require a large number of matched clean images
for training.

To address this issue, researchers have proposed various forms of  self-supervised
learning approaches trained with ingenious forms of  loss functions that are not associated with clean reference images \cite{lehtinen2018noise2noise,krull2019noise2void,batson2019noise2self,huang2021neighbor2neighbor,soltanayev2018training,kim2020unsupervised}. 
Specifically,
these approaches have focused
on designing loss functions to prevent from learning identity mapping, and 
can be categorized into two classes: 1) one with generating altered target images from noisy input images \cite{lehtinen2018noise2noise,krull2019noise2void,batson2019noise2self,huang2021neighbor2neighbor}, 
and  2) the other by adding regularization terms from Stein's Unbiased Risk Estimation (SURE) \cite{soltanayev2018training,kim2020unsupervised}.

Although these algorithms appear seemingly different,
a recent proposal of Noise2Score~\cite{kim2021noise2score} revealed that
the procedure of generating altered target images or SURE-based regularization term  is closely related to the score matching \cite{hyvarinen2005estimation},
and there exists a minimum mean square error (MMSE) optimal denoising formula  in terms of score function for any exponential family
distributions.
Unfortunately, in the case of truly blind image denoising problem where noise statistics  are unknown,
Noise2Score cannot provide an optimal performance. 

%

One of the most important contributions of this paper is, therefore, the discovery that the classic Tweedie distribution can provide
a ``magic'' recipe that can be used for a large class of noise distribution even without knowing the distribution.
Specifically,  Tweedie distribution  can be synergistically combined with Noise2Score to provide an explicit de-noising formulation and an algorithm for estimating the underlying noise distributions and parameters.
In particular, inspired by the fact that various exponential family distributions like Gaussian, Gamma, Poisson, etc. can be described by saddle point approximation of the Tweedie distribution by simply changing one parameter, we provide a universal noise removal formula that can be used for a large class exponential family distributions without prior knowledge of the noise model.
Furthermore, by assuming that slightly perturbed noisy image may produce similar denoising results,
we provide a systematic algorithm that can estimate the noise type and associate parameters for any given images.

In spite of the blind nature of the algorithm,  experimental results demonstrated that
our method outperforms other self-supervised image denoising methods that are trained with prior knowledge of noise distributions.
Our contribution can be summarized as follows.
\begin{itemize}
\item We provide a general closed-form denoising formula for large classes of noise distributions
 by combining Noise2Score approach and the saddle point approximation of Tweedie distribution.
\item We propose an algorithm to estimate the noise model and noise parameter for given noisy images. In particular, the proposed noise estimation algorithm significantly improves the performance and boosts the inference speed compared to the original Noise2Score~\cite{kim2021noise2score}.
\item We show that the proposed method results in the state-of-the-art performance amongst various self-supervised image denoising algorithms in the benchmark dataset and real-environment dataset. 
\end{itemize}

\section{Related Works}

\subsection{Self-supervised  image denoising}
Recently,
 self-supervised image denoising methods using only noisy images have been widely explored.  Noise2Noise~\cite{lehtinen2018noise2noise} (N2N) was proposed to train a neural network by minimizing $L_2$ distance between the noisy image and another noisy realization of the same source image. 
 When additional noisy versions of the same image are not available,
 Noise2Void (N2V)~\cite{krull2019noise2void}, Noise2Self (N2S)~\cite{batson2019noise2self} adopted the blind spot network to avoid learning the identity noisy images. 
Other types of mask-based blind-spot network methods have been explored, which include Self2Self~\cite{quan2020self2self}, Noise2Same~\cite{xie2020noise2same}, Laine19~\cite{laine2019high}, etc.
Instead of using blind-spot methods, 
Neighbor2Neighbor (Nei2Nei)~\cite{huang2021neighbor2neighbor} generate training image pairs by using the random neighbor sub-sampler method. Noisier2Noise~\cite{moran2020noisier2noise} samples doubly noisy images by adding synthetic noises on the single noisy input. 
Additionally, Soltanayev et al ~\cite{soltanayev2018training} proposed a training scheme using Stein’s unbiased risk estimator (SURE) to denoise additive Gaussian noise. By extending this idea, in~\cite{kim2020unsupervised}, the authors proposed a loss function using Poisson Unbiased risk estimator (PURE) to train deep neural network for  Poisson noises. 

However, these methods have been mostly designed either by distribution independent heuristics, or 
 for specific noise models and we are
not aware of any general methods that can be used
for a large class of noise distributions by estimating the noise model and parameters. 

\subsection{Noise2Score}

Recently, 
Noise2Score~\cite{kim2021noise2score} was proposed to estimate clean images by Tweedie’s formula given the learned score function of noisy data. 
%
Specifically, Noise2Score~\cite{kim2021noise2score} consists of two steps:
score function estimation using noisy images, which is followed by
Tweedie's formula to estimate the clean images.
To learn the score function from noisy data $y$, Noise2Score employes the amortized residual DAE (AR-DAE)~\cite{lim2020ar}, which is a stabilized implementation of denoising autoencoder (DAE) \cite{vincent2010stacked}.
In fact, this procedure is closely related to the masking procedure in Noise2Void, Noise2Self, Noiser2Noise, etc.


Once the score function $l'(y)$ for the noisy image is learned, Noise2Score 
provides an explicit formula for the
clean image using Tweedie's formula. More specifically, consider an exponential family of probability distributions:
\begin{eqnarray}
p(y|\eta)=p_0(y)\exp\left(\eta^\top T(y) - A(\eta) \right)
\label{eq:exponen}
\end{eqnarray}
where the superscript $^\top$ denotes the transpose operation. 
Here,  $\eta$ is a canonical (vector) parameter  of the family, $T(y)$ is a (vector) function of $y$,
 $A(\eta)$ is the cumulant generating
function which makes $p(y|\eta)$  integrate to 1, and $p_0(y)$ the density up to a scale factor when $\eta = 0$. 
Then, Tweedie's formula \cite{efron2011tweedie}
shows that the posterior estimate of the canonical parameter $\hat{\eta}$ should satisfy the following
equation:
%
\begin{eqnarray}\label{eq:fix}
\hat\eta^\top  T'(y) &=& -l_0'(y)+l'(y) 
\label{eq:closed}
\end{eqnarray}
where $l'(y):=\nabla_y \log p(y)$ and $l_0'(y):=\nabla_y \log p_0(y)$ are score functions, and $T'(y)=\nabla_y T(y)$ \cite{kim2021noise2score}.


\section{Main Contribution}

In practice, when noisy images are collected, it is difficult to assume the underlying noise models.
For example, low dose X-ray CT images are contaminated with a lot of noises, but it is not clear whether it is from Gaussian noise, Poisson noises, or mixed distribution.
Usually, a specific noise model is assumed for noise removal, but when the assumed noise model is incorrect, there exists an unavoidable bias in the estimated images.
The situation is same in Noise2Score, as a specific  form of Tweedie's formula is used depending on the assumed noise model.
Therefore, a fundamental question for image denoising is whether there is a universal algorithm that can be used for various noise models
without losing its optimality.

In this section, we answer this question affirmatively. Specifically, we show that the classical Tweedie distribution
model is a versatile tool that can be combined with Noise2Score to provide a near optimal solution for various noise distribution
without even knowing the underlying noise model.

%
%
\subsection{Tweedie distribution}

The Tweedie distributions are a family of probability distributions which include the purely continuous normal, gamma and Inverse Gaussian distributions, the purely discrete scaled Poisson distribution, and the class of compound Poisson–gamma distributions \cite{jorgensen1997theory}. 
Specifically, for a random variable $y$ which follows an exponential dispersion model, the Tweedie density function is given by
\begin{eqnarray}
	p(y;\mu,\phi) = b_{\rho}(y,\phi)\exp\left(\frac{-d(y,\mu)}{2\phi}\right),
\end{eqnarray}
where $\mu$ is the mean, $\phi > 0$ is the dispersion parameter which is related to noise level, $d(y, \mu)$ is the unit deviance for $ \rho \neq 0,1,2$ :
\begin{eqnarray}
	d(y,\mu) = 2\left(\frac{y^{2-\rho}}{(1-\rho)(2-\rho)} - \frac{y\mu^{1-\rho}}{1-\rho} + \frac{\mu^{2-\rho}}{2-\rho}\right). \label{deviance}
\end{eqnarray}

Tweedie family densities are characterized by power variance functions of the form $V[\mu]$ = $\phi \mu^{\rho}$, where $\rho \in (-\infty,0] \cup [1,\infty)$ is the index determining the distribution~\cite{jorgensen1997theory}. This has a closed form expression for special cases,  such as the normal distribution ($\rho$ = 0), Poisson ($\rho$ = 1), gamma ($\rho$ = 2) and inverse Gaussian ($\rho$ = 3) distributions (see Table ~\ref{tbl:tweedie}). 
However, the function $b_{\rho}(y,\phi)$ cannot be written in closed form except for  special cases.

\begin{table}[!t]
	\centering
	\caption{Exponential dispersion models with power variance functions. }
	\label{tbl:tweedie}
	\begin{small}	
	\resizebox{0.75\linewidth}{!}{
		\begin{tabular}{cccc}
			\toprule
			{Distribution}	& $\rho$	& $ V[\mu]$	 & $\phi$ \\ \cmidrule(r){1-1} \cmidrule(r){2-4} 
			Gaussian  &  0	& 1 & $\sigma^2$  \\ 
			Poisson	 &	1 & $\mu$	 & 1 \\
			Poisson-Gamma &	1$<\rho<$2 & $\mu^{\rho}$	 & $\phi$ \\
			Gamma($\alpha,\beta)$	 &	2 & $\mu^{2}$	 & $1/\alpha$ \\ 
			Inverse Gaussian &3 & $\mu^{3}$ & $\phi$ \\
			\bottomrule				
	\end{tabular}}
	\end{small}
\end{table} 

Thus, the saddle point approximation can be often used to approximate the Tweedie densities, where
$b_{\rho}(y,\phi)$ is replaced by a simple analytic expression, which leads to a simple expression of the density~\cite{dunn2001tweedie}:
\begin{eqnarray}
p(y;\mu,\phi) = (2\pi\phi y^{\rho})^{-\frac{1}{2}}\exp\left(\frac{-d(y,\mu)}{2\phi}\right).
\label{eq:saddle}
\end{eqnarray}


In the following section, we show that \eqref{eq:saddle} is the key to derive a universal
denoising formula that can be used for various forms of Tweedie distribution.

\subsection{Noise2Score for  Tweedie distribution}
\label{general}

Now using the same idea of Noise2Score that relies on \eqref{eq:closed}, we can obtain the following
universal denoising formula from the saddle point approximation \eqref{eq:saddle}. All the proofs in this section are deferred to Supplementary Material.
%
\begin{prop}
For the given measurement model  \eqref{eq:saddle},
 the MMSE optimal estimate of the unknown $\mu$ is given by
\begin{align}
&\hat{\mu} = \mathbb{E}[\mu|y] = y\left(1+(1-\rho){\alpha(y, \rho,\phi)}\right)^{\frac{1}{1-\rho}} \label{hatx}
\end{align}
where $$\alpha(y, \rho,\phi)= \phi y^{\rho-1}\left(\frac{\rho}{2y} + l'(y)\right).$$
\label{prop}
\end{prop}

\begin{prop}
	The estimate \eqref{hatx}
	converges to the specific  formulae  in Table~\ref{tbl:exp}  for  given parameter pairs $(\rho,\phi)$ .
	\label{corollary}
\end{prop}

\begin{table}[!hbt]
	\centering
	\caption{Special cases of Tweedie's formula for denoising. }
	\label{tbl:exp}
	\vspace{-0.2cm}
	\resizebox{0.7\linewidth}{!}{
			\begin{tabular}{cccc}
				\toprule
				{Distribution}	& $\rho$	&  $\phi$	 & $\hat \mu$	\\ \cmidrule(r){1-1} \cmidrule(r){2-4} 
				Gaussian  &  0	& $\sigma^2$  & $y+\sigma^2 l'(y)$  \\ 
				Poisson	&	1 & $\zeta$	 &$\left(y+\frac{\zeta}{2}\right)\exp(\zeta l'(y))$ \\ 
				Gamma($\alpha,\alpha$) & 2 & $1/\alpha$ & $\frac{\alpha y}{(\alpha-1)-y l'(y)}$\\ 
				\bottomrule				
			\end{tabular}}
\end{table}

Note that the specific formula in Table~\ref{tbl:exp} is equivalent to that of the original paper of Noise2Score \cite{kim2021noise2score}.
This suggests that by simply estimating
the  parameter pair $(\rho,\phi)$, we can estimate both noise model via $\rho$ and 
noise levels with $\phi$. 
In the following, we provide a systematic algorithm to estimate these parameters from a given noisy measurement.

%

\subsection{Noise Model and Level Estimation}

\subsubsection{Noise model estimation}
\label{model}

Here, we provide an algorithm that can be used to estimate the noise model  parameter $\rho$.
Let $y_1$ be the noisy measurement.  Suppose we add a small amount of independent noise to generate $y_1$:
 $$y_2 = y_1 + \epsilon u, \quad u \sim \mathcal{N}(0,I)$$
 where $\epsilon$ is a small known value.
 If the injected noises are sufficiently small, one could expect that
 their denoised images using \eqref{hatx} should be similar, i.e.
 \begin{eqnarray}
 \Ed[\mu~|~y_1]\simeq \Ed[\mu~|~y_2] \label{mu}
 \end{eqnarray}
 Therefore, we have $\alpha(y_1,\rho,\phi)\simeq \alpha(y_2,\rho,\phi)$, which is equivalent to:
 \begin{eqnarray}
\phi y_{1}^{\rho-1}\left(\frac{\rho}{2y_1} + l'(y_1)\right) ~\simeq~ \phi y_{2}^{\rho-1}\left(\frac{\rho}{2y_2} + l'(y_2)\right)\label{alpha2}
\end{eqnarray}

Now, the key observation is that the noise level parameter $\phi$ can be canceled from \eqref{alpha2} when the equality holds.
Therefore, \eqref{alpha2} provides a closed form formula for the noise model parameter $\rho$ as stated in the following proposition.
%
%

\begin{prop}
Suppose that equality holds in \eqref{alpha2} for $y_1\neq y_2$. 
Then, the distribution parameter  $\rho$ is given by
\begin{eqnarray}
		\hat{\rho} = \frac{-a(b-2) \pm \sqrt{(a(b-2)^2 -4a(-2ab +w)}}{2a}
\end{eqnarray}
where $a = \log(\frac{y_2}{y_1}), b = 2y_1l'(y_1)$ and $w = 2y_2 l'(y_2) - 2y_1 l'(y_1)$.
\label{rho}
\end{prop}

Proposition $\ref{rho}$ implies that we can estimate the unknown noise model by solving the quadratic equation. 
Among two solutions for $\rho$, we empirically determine $\hat{\rho}$ by taking the maximum of two values for the correct estimate of $\rho$. Since we assume that $\rho \in (0,2)$, we set the final $\hat{\rho}$ to $\max(0,\hat{\rho})$.

To determine whether underlying noises are either from Gaussian, Poisson, and Gamma,
 we set the empirical rule to estimate the noise model from  $\hat \rho$:
\begin{align}
 \begin{cases*}		
        \text{Gaussian}, & if $ 0 \leq \hat{\rho} < 0.9 $,\\
        \text{Poisson}, & if $ 0.9 \leq \hat{\rho} < 1.9 $,\\
        \text{Gamma}, & if $ 1.9 \leq \hat{\rho} < 2.9 $.
        \end{cases*}
\label{rule}        
\end{align}
Specifically, if the estimated model parameter $\hat{\rho}$ is included in a specific range, we conclude that the noisy images belong to corresponding the noise distribution. 
The detail of implementation are described in Supplementary Material.

\subsubsection{Noise level estimation}
\label{level}

If the noise model parameter $\rho$ is known or correctly estimated, 
the universal denoising formula in \eqref{hatx} still require the knowledge of noise level $\phi$.

In contrast to the original Noise2Score, which estimates noise level using image quality penalty metric such as total variation (TV), 
here we propose a novel method to estimate the unknown noise parameter much more efficiently. 
Again, the basic intuition is that
 if the injected noises are sufficiently small, one could expect that
 their denoised images using \eqref{hatx} should be similar.
Proposition~\ref{noiselevel} provides the specific formula based on the assumption.
\begin{prop}
Assume that the noise model parameter $\rho$ is known. Suppose, furthermore,
that  the equality holds in \eqref{mu} for $y_1\neq y_2$. Then, 
the noise level parameter of several noise distribution is given by
	\begin{align}
	\begin{cases}
	 \hat{\sigma}^2 = \left(\frac{-\epsilon u}{l'(y_2) -   l'(y_1)}\right),&\mbox{Gaussian},\\
	\hat{\zeta} = \left(-y_1 + \sqrt{y_1^2 - 2c}\right),&\mbox{Poisson}, \;\\
	\hat{\alpha} = \hat{\beta} = \hat{k} = \left(1 + \frac{l'(y_2) - l'(y_1)}{\frac{1}{y_2} - \frac{1}{y_1}}\right), &\mbox{Gamma}, \;
		 \end{cases}	
	\end{align}	
where $c = \epsilon u/(l'(y_2)-l'(y_1))$.
\label{noiselevel}
\end{prop}
For the case of Poisson noise, the solution is derived from the quadratic equation for $\zeta$. Due to $\zeta >$ 0, we only calculate the solution with a positive sign.  
All  these formulae are applied for each pixel level, so  we should use the median for the pixel level estimate. 

In contrast to Noise2Score which requires multiple inferences for the noise level  estimation, our proposed method requires  only one more inference step to find unknown noise level parameter. Thus, the proposed method has advantages in terms of speed and simplicity.

\begin{figure*}[ht!]
	\centering
	\includegraphics[width=1\linewidth]{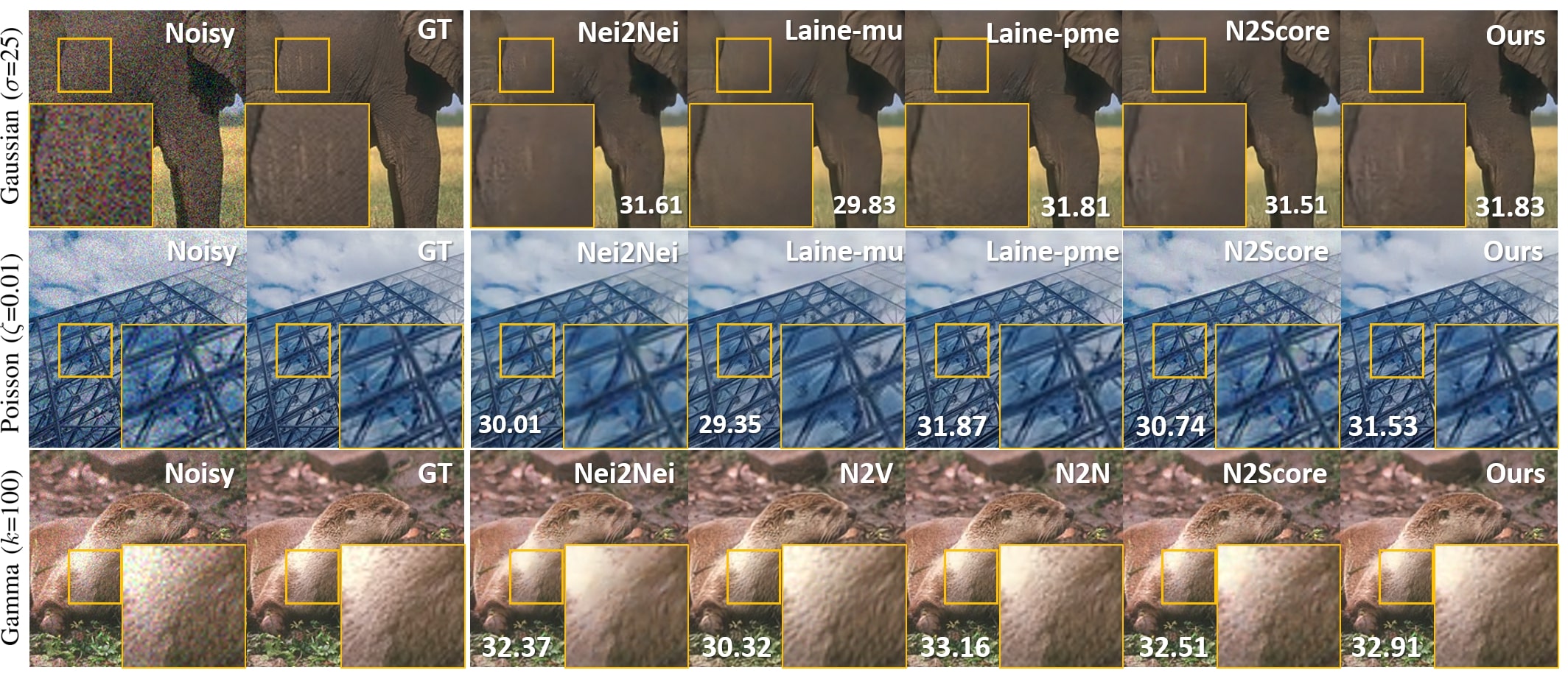} 
	\caption{Visual comparison of our method against other competing methods in CBSD68 dataset. Top : Gaussian noise with $\sigma$ = 25. Middle: Poisson noise with $\zeta$ = 0.01. Bottom: Gamma noise with $k$ = 100. The number at the lower corner of images indicates the PSNR value in dB.}
	\label{fig:result1}
\end{figure*}
\begin{figure*}[ht!]
	\centering
	\includegraphics[width=1\linewidth]{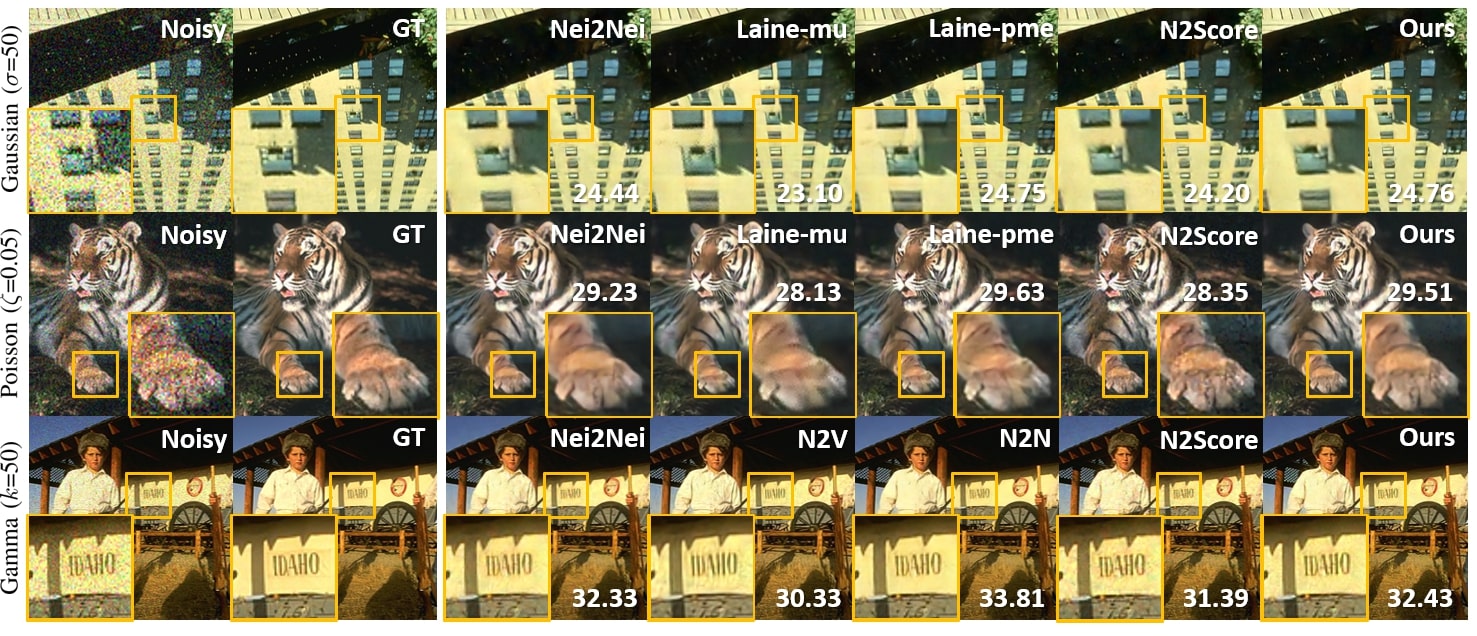} 
	\caption{Visual comparison of our method against other competing methods in CBSD68 dataset. Top : Gaussian noise with $\sigma$ = 50. Middle: Poisson noise with $\zeta$ = 0.05. Bottom: Gamma noise with $k$ = 50. The number at the lower corner of the image indicates the PSNR value in dB.}
	\label{fig:result2}
\end{figure*}
\begin{table*}[t!]
	\begin{small}	
\begin{center}
			\caption{Quantitative comparison of denoising results by various methods in terms of PSNR(dB) when the noise parameters are unknown 
			(N2V: Noise2Void, N2S: Noise2Self, Nei2Nei: Neighbor2Neighbor, N2Score: Noise2Score, N2N: Noise2Noise, SL: supervised learning, Anscombe: Anscombe method for BM3D).}
				\label{tbl-results}
\resizebox{0.92\textwidth}{!}
{%
\begin{tabular}{clcccccccccc}\toprule
	\multicolumn{1}{c}{Noise type}                    & & \multicolumn{1}{c}{Conventional}                                     & \multicolumn{7}{c}{Self-Supervised} & \multicolumn{2}{c}{Supervised}\\ \cmidrule(r){3-3}\cmidrule(r){4-10} \cmidrule(r){11-12}
	Gaussian                    & Dataset & BM3D     & N2V   & N2S   & Nei2Nei  & Laine19-mu & Lain19-pme & N2Score   & Ours   & N2N   & SL   \\ \cmidrule(r){1-2} \cmidrule(r){3-3}\cmidrule(r){4-10} \cmidrule(r){11-12}
	\multirow{2}{*}{$\sigma=$ 25} 
	& CBSD68  & 28.61    & 29.22 & 29.71 & 30.56 & 28.61 & 30.88 & 30.75 & \textbf{30.89}  & 30.92   & 30.92    \\
	& Kodak   & 29.94   & 30.02 & 30.81 & 31.55 & 30.19 & 31.92 &  31. 78 & \textbf{31.95} & 31.96   & 31.96     \\ \cmidrule(r){1-2} \cmidrule(r){3-3} \cmidrule(r){4-10} \cmidrule(r){11-12}
	\multirow{2}{*}{$\sigma$ = 50}  
	& CBSD68  & 26.71    & 25.13 & 27.14 & 27.32 & 26.42 & \textbf{27.65} & 27.32 & 27.56  & 27.73  & 27.73     \\
	& Kodak   & 27.02    & 25.75 & 28.21 & 28.28 & 27.78 & 28.63 & 28.23 & \textbf{28.64}  & 28.70  & 28.71     \\ \cmidrule(r){1-2} \cmidrule(r){3-3}\cmidrule(r){4-10} \cmidrule(r){11-12}
	Poisson                    & Dataset & Anscombe & N2V   & N2S   &  Nei2Nei &Laine19-mu & Lain19-pme  & N2Score & Ours & N2N   & SL   \\ \cmidrule(r){1-2} \cmidrule(r){3-3}\cmidrule(r){4-10} \cmidrule(r){11-12}
	\multirow{2}{*}{$\zeta$ = 0.01} 
	& CBSD68  & 30.68    & 31.02 & 30.74 & 31.64 & 30.61 & \textbf{32.73} & 31.87 & 32.53   & 32.94  & 32.95     \\
	& Kodak   & 31.93    & 31.98 & 31.92 & 32.61 & 31.67 & \textbf{33.51} & 32.96 & 33.41   & 33.86  & 33.87     \\ \cmidrule(r){1-2} \cmidrule(r){3-3} \cmidrule(r){4-10} \cmidrule(r){11-12}
	\multirow{2}{*}{$\zeta$ = 0.05} 
	& CBSD68  & 26.93    & 28.12 & 28.33 & 28.46 & 27.53 & \textbf{28.83} & 28.27 & 28.73  & 29.07  & 29.07     \\
	& Kodak   & 28.27    & 29.32 & 29.49 & 29.54 & 28.68 & \textbf{30.17} & 28.98 & 29.72  & 30.23  & 30.25     \\ \cmidrule(r){1-2} \cmidrule(r){3-3} \cmidrule(r){4-10} \cmidrule(r){11-12}
	Gamma                   & Dataset &       & N2V   & N2S   &  Nei2Nei &  &  & N2Score & Ours           & N2N   & SL   \\ \cmidrule(r){1-2} \cmidrule(r){3-3}\cmidrule(r){4-10} \cmidrule(r){11-12}
	\multirow{2}{*}{$k$ = 100}
	& CBSD68  & -        & 31.83 & 31.71 & 34.21 & - & - & 33.82 & \textbf{34.52} & 35.33  & 35.33     \\
	& Kodak   & -        & 31.66 & 32.83 & 35.10     & - & - & 34.23 & \textbf{35.42}  & 36.16   & 36.16    \\ \cmidrule(r){1-2} \cmidrule(r){3-3} \cmidrule(r){4-10} \cmidrule(r){11-12}
	\multirow{2}{*}{$k$ = 50}
	& CBSD68  & -        & 30.51 & 30.63 & 32.11     & - & - & 31.32 & \textbf{32.43}  & 33.41  & 33.41     \\
	& Kodak   & -        & 31.38 & 31.71 & 32.38     & - & - & 31.81 & \textbf{32.81}  & 34.39  & 34.40   \\ \bottomrule
\end{tabular}}
	\end{center}
	\end{small}
\vspace{-1em}
\end{table*}

\section{Experimental Results}
\subsection{ Implementation Details}
\paragraph{Training Details} In order to fairly compare the proposed method with other comparison methods, a modified version of U-Net generator~\cite{krull2019noise2void} is used for all methods. The mini-batch size was adopted as 16 and the total epoch was set to 100 for training. The Adam optimizer~\cite{kingma2014adam} was used to train the neural network with an initial learning rate 2$\times 10^{-4}$. The learning rate was decayed to $2\times 10^{-5}$ after 50 epochs. To train the score model in this work, we employed the AR-DAE in Noise2Score method~\cite{kim2021noise2score} and adopt the exponential moving average method with decay rate 0.999. For the annealing sigma $\sigma_a$  which is required to learn the score function of noisy data, we generate geometric sequences which have the value of the perturbed noise level from $\sigma_a^{max}$ to $\sigma_a^{\min}$~\cite{song2020improved}.
Further details of implementation are described in Supplementary Material.
Our method was implemented in PyTorch~\cite{paszke2017automatic} with NVIDIA GeForce GTX 2080-Ti. 

\paragraph{Datasets for Synthetic Experiments} We used DIV2K~\cite{Timofte_2018_CVPR_Workshops} and CBSD400~\cite{martin2001database} dataset as  training data set. We sampled 220,000 cropped patches of 128$\times$128 size to train the network. For the data augmentation, we used the random horizontal, vertical flip, and flop methods. 
We generated the synthetic noise images for blind noise cases where noise levels are randomly sampled from the particular range for each noise distribution. In the case of Gaussian noise, noise levels varied with $\sigma \in$ [5,55]. For Poisson noise, noise levels varied with $\zeta \in$ [0.1,0.005]. For Gamma noise, noise levels are in $k \in [40,120]$. 
To evaluate the proposed method for blind synthetic noise experiment, the test set was adopted for Kodak, CBSD68~\cite{martin2001database} dataset.

\subsection{Results of Synthetic Experiments.}
\paragraph{Gaussian noise}
To evaluate our method on various noise distribution, we adopted the nine comparison methods such as BM3D~\cite{dabov2006image}, N2V~\cite{krull2019noise2void}, N2S~\cite{batson2019noise2self}, Nei2Nei~\cite{huang2021neighbor2neighbor}, Laine19~\cite{laine2019high}, N2Score~\cite{kim2021noise2score}, N2N~\cite{lehtinen2018noise2noise}, and supervised learning approaches as shown in Table \ref{tbl-results}. Comparison methods are varied depending on the noise distribution. The supervised learning and N2N using multiple pair images perform best, but are not practical. In the case of additive Gaussian noise, our method not only outperforms the other self-supervised learning approaches for all of dataset but also provides comparable results to the supervised learning approaches. 
Thanks to the improved score function estimation described in Supplementary Material, our method provides even better performance of N2Score. The qualitative comparison in Figs.~\ref{fig:result1} and \ref{fig:result2} show that the proposed method provides the best reconstruction results. 

\vspace{-1em}
\paragraph{Poisson noise}
For the Poisson noise case, BM3D was replaced with Anscombe transformation (BM3D+VST) as indicated in Table~\ref{tbl-results}. Our proposed method provides significant gain in performance compared to other algorithms, although Laine19-pme, taking advantage of the noise model with known prior, shows the best performance for the Poisson case. Note that our proposed method assumes that the noise model is unknown and estimates the noise statistics. Despite the unknown noise model, the results of proposed method are still comparable. We found that our method significantly improved the results of Noise2Score in Poisson case when the noise level parameter is unknown. It confirmed that the proposed noise level estimation method is more effective than quality penalty metrics in Noise2Score. 
The visual comparison results in Figs.~\ref{fig:result1} and \ref{fig:result2} show that our method delivers much visually pleasing results compared to other self-supervised methods. 
\vspace{-1em}
\paragraph{Gamma noise}
As the extension of BM3D and Laine19 for Gamma noises are not available, we only used six comparison methods to evaluate the proposed method, as indicated in Table~\ref{tbl-results}. In the case of Gamma noise case, we set $\alpha =\beta = k$. Again, the proposed method yields the best results against self-supervised learning approaches and significantly improves the performance of Noise2Score with a margin of $\sim$1dB. The qualitative comparison in Figs.~\ref{fig:result1} and \ref{fig:result2} confirm that our method provides competitive visual quality among self-supervised deep denoisers. 
 
\begin{figure*}[!t]
 	\centering
 	\includegraphics[width=0.95\linewidth]{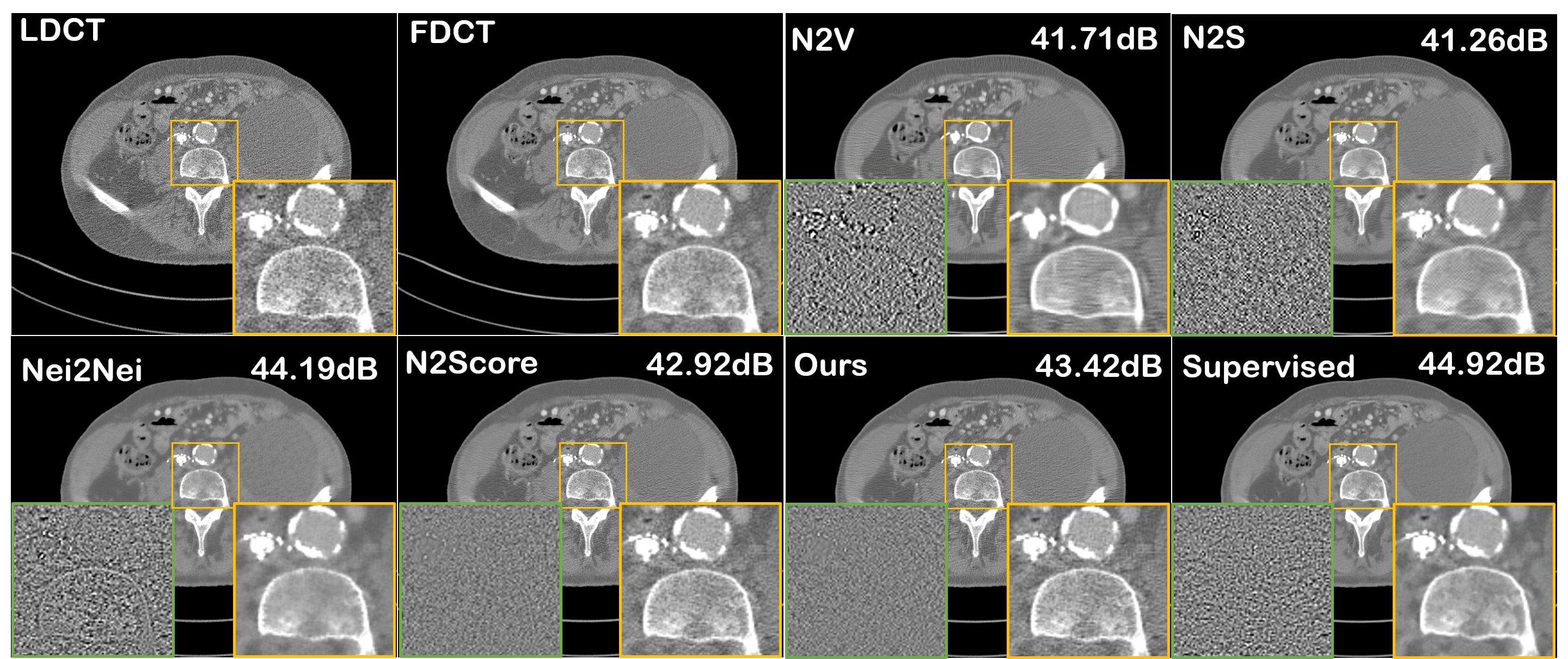} 
 	\caption{Denoising results of AAPM data using various methods. The yellow box and green box show the enlarged view of image and difference image between network input and output, respectively. The intensity window of CT image is (-500,500)[HU] and the intensity window of difference is (-200,200) [HU].}
 	\label{fig:ct}
\end{figure*}

\subsection{Real Image Noise Removal}
To verify that the algorithm can be applied to a real dataset, a real noise removal experiment was performed using the AAPM CT dataset~\cite{mccollough2017low}, which contains 4358 images in 512$\times$512 resolution from 10 different patients at low-dose level and high-dose X-ray levels. The input images are adopted as quarter dose images and the target images are set to full dose images. We randomly select 3937 images as a train set, and the remaining 421 images are set as a test set. The noise distribution in X-ray photon measurement and in the sinogram are often modeled as Poisson noise and Gaussian noise, but real noise in reconstructed images is spatially correlated and more complicated, so that it leads to difficulties in modeling the specific noise distributions. 

To deal with blind image noise removal in the low dose CT images, we carried out experiment with the proposed method using the procedure in Fig. \ref{fig:concept}. The other experiment settings are identical to synthetic experiments. Based on the model estimation rule in \eqref{rule}, we observed that the Low-dose CT images can be interpreted as corrupted by Gaussian noise ($\hat{\rho}\simeq$ 0). To evaluate the proposed method, we compare with other methods such as N2S~\cite{batson2019noise2self}, N2V~\cite{krull2019noise2void}, Nei2Nei~\cite{huang2021neighbor2neighbor}, N2Score~\cite{kim2021noise2score}, and supervised learning approach. To implement the original Noise2Score, we assume that the noise distribution is additive Gaussian noise. The quantitative results suggest that the proposed method yields the competitive results to other self-supervised learning approaches. However, the visual quality is also an important criterion to evaluate the results in the medical imaging. In Fig \ref{fig:ct}, the qualitative comparison shows that other self-supervised approaches provide over-smooth denoised images. In particular, the difference images of these methods show the structure of CT images. However, our method yields visually similar results compared to full-dose CT image and also provides  only noises in the difference image between the network input and output. Therefore, our method also has  the advantage that it can be applied to real environment dataset.

\begin{table}[h!]
	\begin{small}		
		\begin{center}
			\caption{Quantitative comparison using various methods in terms of PNSR(dB) on the AAPM CT dataset. (N2V: Noise2Void, N2S: Noise2Self, Nei2Nei: Neighbor2Neighbor, N2Score: Noise2Score  SL: supervised learning).}
			\label{tbl:CT}
			\resizebox{0.95\linewidth}{!}{
				\begin{tabular}{ccccccc}\toprule
					Method  & N2S & N2V  & Nei2Nei & N2Score&Ours &SL \\ \cmidrule(r){1-1} \cmidrule(r){2-6} \cmidrule(r){7-7}
					PSNR  & 35.82 & 35.93  & 37.83 & 36.63 & 36.92 &38.53\\ \bottomrule
			\end{tabular}}
		\end{center}
	\end{small}
\vspace{-1em}
\end{table}

\section{Ablation Study}
\subsection{Results on Noise model estimation}
\begin{figure}[!t]
	\centering
	\begin{subfigure}{0.494\linewidth}
		\centering
		\includegraphics[width=1\linewidth]{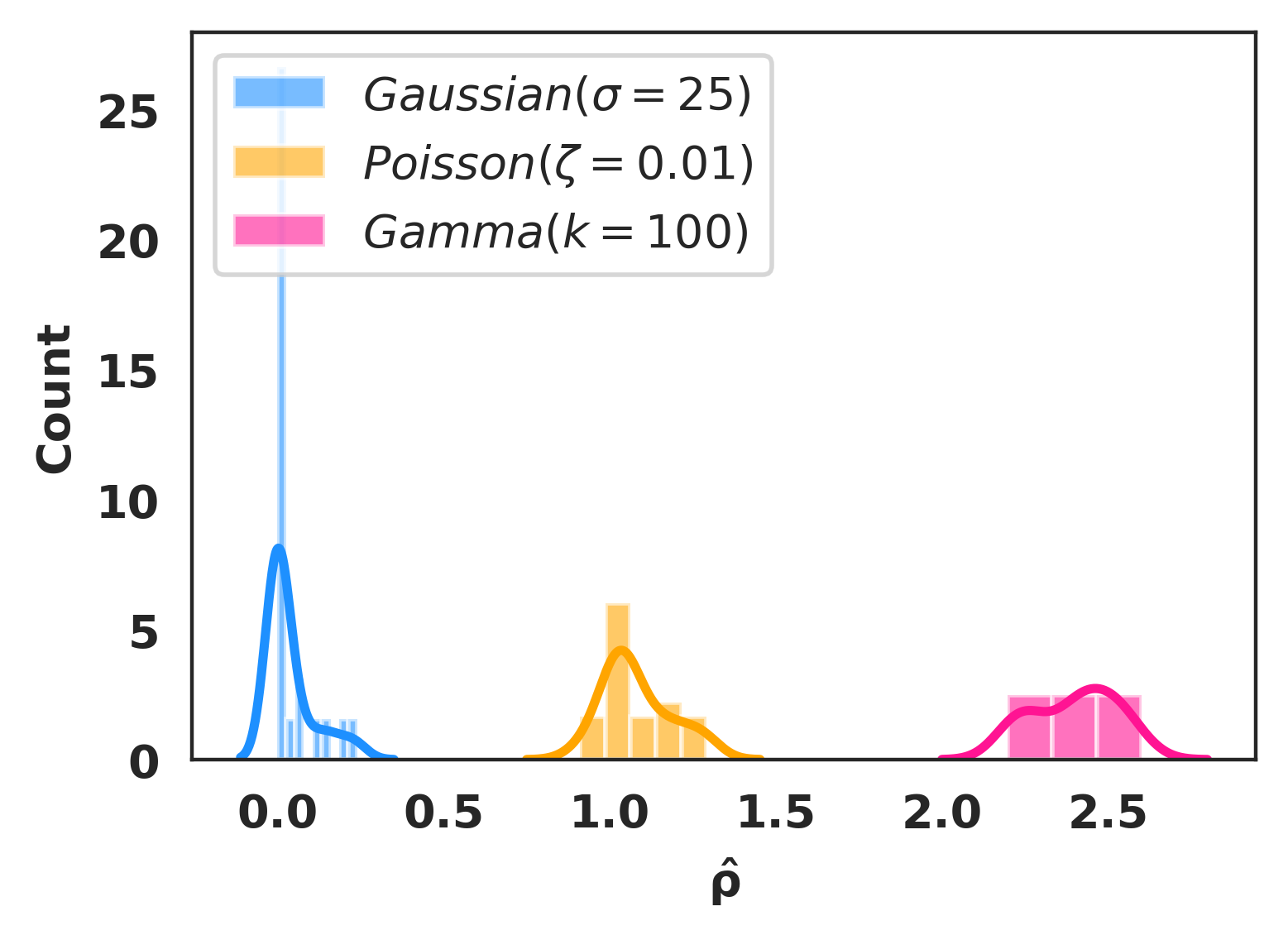} 
		\caption{} 
	\end{subfigure}
	\begin{subfigure}{0.494\linewidth}
		\centering
		\includegraphics[width=1\linewidth]{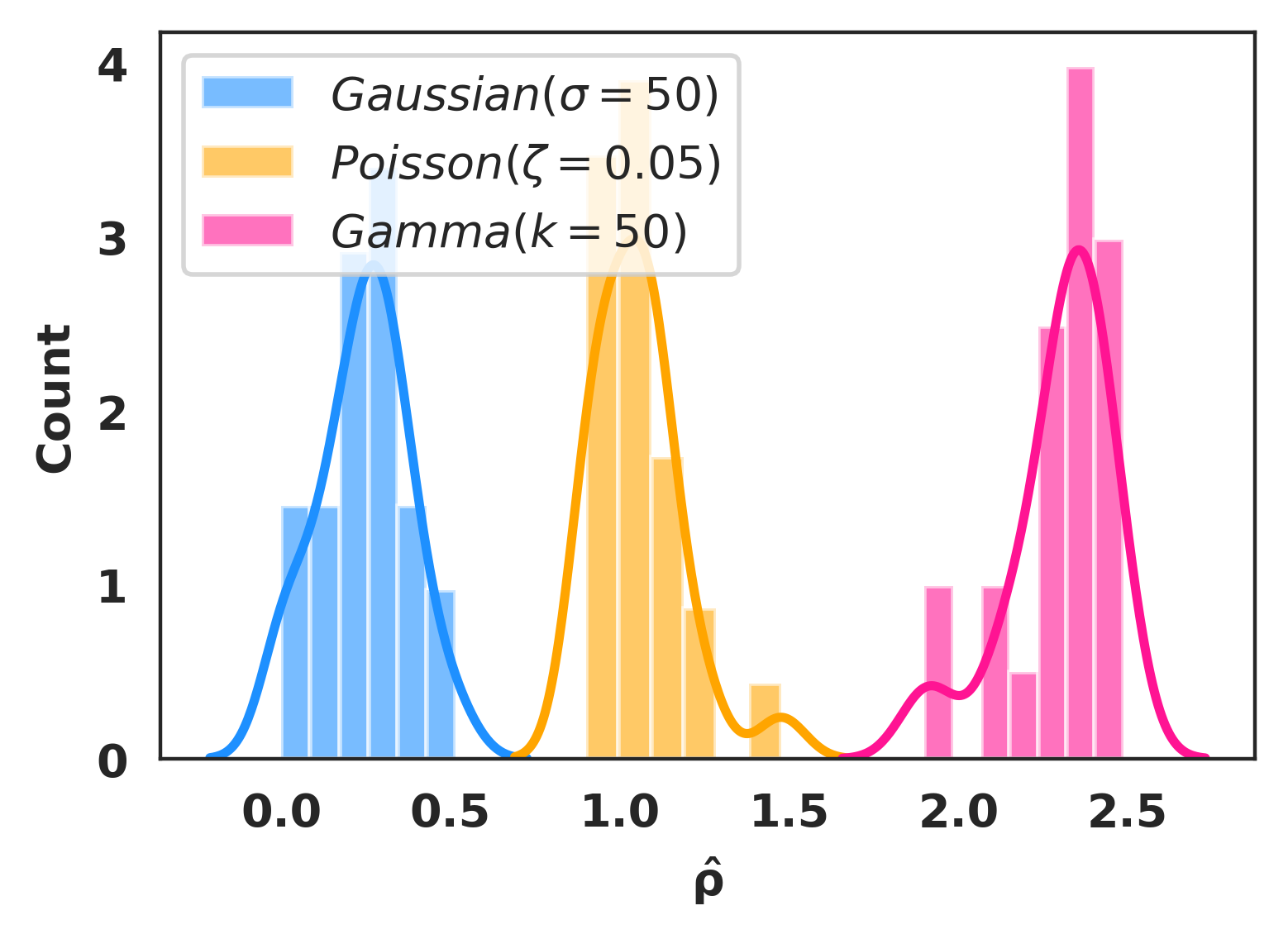}
		\caption{} 
	\end{subfigure}
	\vspace{-1em}
	\caption{The histogram of estimated noise model parameter $\hat{\rho}$ in Kodak dataset. (a) For the case of $\sigma$ =25, $\zeta$ = 0.01, $k$ = 100.  (b) For the case of $\sigma $=50, $\zeta$ = 0.05, $k$ = 50.}
	\label{fig:histo-noise}
	\vspace{-1em}
\end{figure}

Here, we validate the proposed the noise model estimation by \eqref{rule}. Figure~\ref{fig:histo-noise} shows the histogram of estimated noise model parameter $\hat{\rho}$ in the Kodak dataset. We have calculated $\hat{\rho}$ for each noise distribution and plot distributions in each figure. According to Tweedie's distribution, we expect that the parameters of noise model can be differentiated for each noise distribution such as Gaussian ($\hat{\rho}$ $\simeq$ 0), Poisson ($\hat{\rho}$ $\simeq$ 1), and Gamma ($\hat{\rho}$ $\simeq$ 2). From the figure, we can observe that the noise model parameters $\hat{\rho}$ are distinctly distributed in the case of low noise levels. Furthermore, in the case of high noise levels, the distributions can also be distinguished for each noise distribution. It implies that it was possible to successfully estimate the noise model with the proposed method, so that it leads to the use of the exact Tweedie's formula for blind image denoising. 

\begin{table}[ht!]
	\begin{small}
		\begin{center}
			\caption{Reconstruction results with various noise parameter estimation methods on CBSD68 dataset in terms of PSNR(dB).}
			\vspace{-1em}
			\label{tbl:ablation}
			\resizebox{0.9\linewidth}{!}{
				\begin{tabular}{ccccc}\toprule
					Noise type  & Quality Penalty & Ours  & Known value & SL \\ \cmidrule(r){1-1} \cmidrule(r){2-4} \cmidrule(r){5-5}
					Gaussian ($\sigma$=25)  & 30.78 & 30.89  & 30.91 & 30.92    \\
					Poisson ($\zeta$=0.01)  & 31.89 & 32.53  & 32.63 & 32.95    \\
					Gamma ($k$=100)         & 33.92 & 34.52  & 34.53 & 35.33    \\\midrule
					Inference speed & 3.1s & 0.1s & -&- \\ \bottomrule
			\end{tabular}}
		\end{center}
	\end{small}
	\vspace{-2em}
\end{table}

\subsection{Ablation study on Noise level estimation}
We analyze the effect of the proposed noise level estimation by comparing quality penalty metric\cite{kim2021noise2score} and others. To fairly evaluate the proposed method, we used an identical weights of the score model for each case. We carried out an ablation study by fixing the noise model estimation procedure and by only varying the noise level estimation as shown Table~\ref{tbl:ablation}. Here,  ``known value" refers to the results assuming the known ground truth noise level parameter. ``SL" denotes that the results of supervised learning. We found that quality penalty metric yield comparable performance to ``known value" in the case of the Gaussian, but the performance with the quality penalty metric decreased for the case of Poisson and Gamma noises. However, the proposed methods outperform the quality penalty metric approach and yields results comparable to ``known value" in all  cases. Furthermore, we compare the inference speed of the estimation stage. The quality penalty metric method took about $\times$ 30 times more to estimate noise level parameters compared to our method.

\section{Conclusion}
In this article, we provided a novel self-supervised blind image denoising framework that does not require clean data and  prior knowledge of
 noise models and levels.  Our innovation came from the saddle point approximation of Tweedie distributions, which cover a wide range of exponential
 family distributions. By taking advantage of this property, we provided a universal denoising formula
 that can be used for various distributions in real life. Furthermore, we proposed a novel algorithm that can estimate the noise model and noise level parameter in a unified framework.  Finally, we validated the proposed method using  benchmark and real CT image data sets,
 and confirmed that the method outperforms the existing state-of-the-art self-supervised learning methods.

\appendix
\section*{Appendix}
\section{Proof of Proposition 1}
\label{general}
\begin{proof}
	For a given exponential family of probability distributions:
	\begin{eqnarray}
		p(y|\mu)=p_0(y)\exp\left(\mu^\top T(y) - A(\mu) \right) \ ,
		\label{eq:exponen}
	\end{eqnarray}
	Tweedie’s formula~\cite{efron2011tweedie}  shows that the posterior estimate of the canonical parameter $\hat{\mu}$ should satisfy the following equation:
	\begin{eqnarray}\label{eq:fix}
		\hat\mu^\top  T'(y) &=& -l_0'(y)+l'(y) 
		\label{eq:closed}
	\end{eqnarray}
	where $l'(y):=\nabla_y \log p(y)$ and $l_0'(y):=\nabla_y \log p_0(y)$ are score functions, and $T'(y)=\nabla_y T(y)$ \cite{kim2021noise2score}.

	Now, our goal is to use this formula to the saddle point approximation of Tweedie distribution\cite{dunn2001tweedie} given by:
	\begin{eqnarray}
		p(y;\mu,\phi) = (2\pi\phi y^{\rho})^{-\frac{1}{2}}\exp\left(\frac{-d(y,\mu)}{2\phi}\right)
		\label{eq:saddle}
	\end{eqnarray}
	where 
	\begin{eqnarray}
		d(y,\mu) = 2\left(\frac{y^{2-\rho}}{(1-\rho)(2-\rho)} - \frac{y\mu^{1-\rho}}{1-\rho} + \frac{\mu^{2-\rho}}{2-\rho}\right).
	\end{eqnarray}
	By inspection of  \eqref{eq:exponen} and \eqref{eq:saddle}, we have
	\begin{align}
		&p_{0}(y) = (2\pi\phi y^{\rho})^{-\frac{1}{2}} \nonumber\\
		&\mu^\top T(y) - A(\mu)  =  \frac{-d(y,\hat\mu)}{2 \phi} \label{eq:same}
	\end{align}
	Furthermore, we have
	\begin{align*}
		\frac{\partial d(y,\mu)}{\partial y} &= \frac{2}{1-\rho}y^{1-\rho} - \frac{2}{1-\rho}\mu^{1-\rho} \\
		\frac{\partial \log(2\pi\phi y^\rho)^{-\frac{1}{2}}}{\partial y} &= -\frac{\rho}{2y}
	\end{align*}
	Accordingly,
	\begin{align*}
		\hat\mu^\top  T'(y)&= -\frac{1}{\phi(1-\rho)}(y^{1-\rho}-\hat\mu^{1-\rho}) \\
		&= \frac{\rho}{2y} + l'(y)
	\end{align*}
	which leads to
	\begin{align}
		& \hat\mu^{1-\rho} = y^{1-\rho} + \phi(1-\rho)\left(\frac{\rho}{2y} + l'(y)\right).
		\label{ge:init}
	\end{align}
	Therefore, we have
	\begin{align}
		\hat\mu &= \exp \left\{ \frac{1}{1-\rho}\log\left( y^{1-\rho} + \phi(1-\rho)\left(\frac{\rho}{2y} + l'(y)\right)\right)\right\} \notag \\
		&= y \left(1+ (1-\rho)\alpha(y,\rho,\phi)\right)^{\frac{1}{1-\rho}}\label{eq:mu}
	\end{align}
	%
	where 
	\begin{align}\label{eq:alpha}
		\alpha(y, \rho,\phi)= \phi y^{\rho-1}\left(\frac{\rho}{2y} + l'(y)\right).
	\end{align}
	This concludes the proof.
\end{proof}
\section{Proof of Proposition 2}
\noindent {\em Proof:}
\paragraph{Additive Gaussian noise.} In this case, we have $\rho = 0, \phi = \sigma^2$ for Tweedie distribution.
Accordingly, \eqref{eq:alpha} can be simplified as
\begin{align}
	\alpha(y,0,\phi) = \sigma^2 y^{-1}l'(y) \label{alpha-gau}
\end{align}
Therefore, using \eqref{eq:mu}, we have
\begin{align*}
	\hat{\mu} & = y(1+\sigma^2 y^{-1}l'(y)) = y + \sigma^2 l'(y).
\end{align*}

\paragraph{Poisson noise.} In this case, we have $\rho = 1, \phi = \zeta$ for Tweedie distribution.
In this case, we have
\begin{align}
	& \lim_{\rho \rightarrow 1}(1+ (1-\rho)\alpha(y,\rho,\phi))^{\frac{1}{1-\rho}} \nonumber\\
	=&\; \exp\left[\lim_{\rho \rightarrow 1}\frac{
		\log(1+ (1-\rho)\alpha(y,\rho,\phi))}{1-\rho}\right]\nonumber\\
	=&\; \exp\left[\lim_{\rho \rightarrow 1}\frac{
		\alpha(y,\rho,\phi)}{1+(1-\rho)\alpha(y,\rho,\phi)}\right] \nonumber\\
	=&\; \exp[\alpha(y,1,\phi)] .\label{poi-closed}
\end{align}
where the second equality comes from the L'Hospital's rule,
and
\begin{align}
	\alpha(y,1,\phi) = \zeta \left(\frac{1}{2y} +l'(y)\right) \label{alpha-poi}
\end{align}
Therefore,  we have
\begin{align*}
	& \hat{\mu} = y \exp\left[\zeta\left(\frac{1}{2y}+l'(y)\right)\right]\\
	& = y \exp(\frac{\zeta}{2y}) \exp(\zeta l'(y)) \\
	& \approx (y+\frac{\zeta}{2})\exp(\zeta l'(y)) ,
\end{align*}
where the last approximation comes from $\exp(x) \approx 1+x$ for a small $x$.

\paragraph{Gamma noise.} In this case, we have $\rho =2, \phi = 1/k$ for Tweedie distribution.
Using \eqref{eq:alpha}, we have
\begin{align}
	\alpha(y,2,k) = \frac{1}{k} y\left(\frac{1}{y} +l'(y)\right) = \frac{1}{k}(1+yl'(y)) \label{alpha-gamma}
\end{align}
Therefore, we have
\begin{align*}
	\hat{\mu} & = y\left(1-\frac{1}{k}(1+y l'(y))\right)^{-1} \\
	& = \frac{ky}{(k -1) - yl'(y))} 
\end{align*}	
This concludes the proof.

\section{Proof of Proposition 3}
\begin{proof}
	Let $y_2 = y_1 + \epsilon u$ for $u\sim \mathcal{N}(0,1)$. Then, 
	we have
	\begin{eqnarray}
		\alpha(y_1,\rho,\phi) &=& \phi y_{1}^{\rho-1}\left(\frac{\rho}{2y_1} + l'(y_1)\right) \label{alpha1}\\
		\alpha(y_2,\rho,\phi) &=& \phi y_{2}^{\rho-1}\left(\frac{\rho}{2y_2} + l'(y_2)\right)\label{alpha2}
	\end{eqnarray}
	For a sufficiently small perturbation $\epsilon$, we can assume that
	$$\alpha(y_1,\rho,\phi) = \alpha(y_1,\rho,\phi) $$
	Accordingly, by dividing ($\ref{alpha2}$) by ($\ref{alpha1}$), we have
	\begin{eqnarray}
		1 &=& \left(\frac{y_2}{y_1}\right)^{\rho-1}\left(\frac{\frac{\rho}{2y_2}+l'(y_2)}{\frac{\rho}{2y_1}+l'(y_1)}\right) \nonumber
	\end{eqnarray}
	By taking logarithm on both sides, we have 
	\begin{eqnarray}
		(\rho-2)\log\left(\frac{y_2}{y_1}\right) +\log\left(\frac{\rho+2y_2 l'(y_2)}{\rho+2y_1 l'(y_1)}\right) =0
		\label{eq:19}
	\end{eqnarray}
	Furthermore, by denoting $w:=2y_2 l'(y_2) - 2y_1 l'(y_1) $, we have
	\begin{align}
		\log\left(\frac{\rho+2y_2 l'(y_2)}{\rho+2y_1 l'(y_1)}\right)
		&= \log\left(1+ \frac{w}{\rho+2y_1 l'(y_1)}\right)  \nonumber \\
		&\approx \frac{w}{p+2y_1 l'(y_1)} 
		\label{eq:appro}
	\end{align}
	when $\frac{w}{\rho+2y_1 l'(y_1)} \rightarrow 0.$
	By plugging (\ref{eq:appro}) into (\ref{eq:19}),
	we can obtain the quadratic equation for $\rho$:
	\begin{align}
		& 0 = (\rho-2)\log\left(\frac{y_2}{y_1}\right) +\frac{2y_2 l'(y_2) - 2y_1 l'(y_1)}{\rho+2y_1 l'(y_1)} \nonumber
	\end{align}
	By denoting $a = \log(\frac{y_2}{y_1}), b = 2y_1l'(y_1)$, we have
	\begin{align}
		0 & = a(\rho-2) +\frac{w}{\rho+b}, \nonumber \\
		& = a(\rho -2)(\rho+b) + w. \label{quad}
	\end{align}
	Therefore, the estimated noise model parameter $\hat{\rho}$ can be obtained as the solutions for the quadratic equation, which is given by:
	\begin{eqnarray}
		\hat{\rho} = \frac{-a(b-2) \pm \sqrt{(a(b-2)^2 -4a(-2ab +w)}}{2a} \nonumber
	\end{eqnarray}
\end{proof}

\section{Proof of Proposition 4}
\begin{proof}
	Let $y_2 = y_1 + \epsilon u$ for $u\sim \mathcal{N}(0,1)$, and the noise model parameter $\rho$ is known.
	Suppose furthermore that $\epsilon$ is non-zero but sufficiently small that the following equality holds:
	\begin{align}
		\Ed[\mu|y_1] = \Ed[\mu|y_2] \label{mu} 
	\end{align}
	Now, we derive the formula for each distribution.
	\item
	\paragraph{Additive Gaussian noise}
	For the case of additive Gaussian noise, we have 
	\begin{align}
		& \hat{x}_1 = \Ed[\mu|y_1] = y_1 + \sigma^2l'(y_1) \label{gau_1} \\
		& \hat{x}_2 = \Ed[\mu|y_2] = y_2 + \sigma^2l'(y_2) \label{gau_2}
	\end{align}
	By subtracting \eqref{gau_1} from \eqref{gau_2}, we have
	\begin{align}
		-\epsilon u &= \sigma^2(l'(y_2) - l'(y_1))
	\end{align}
	Thus, we have the following estimate:
	\begin{eqnarray}
		\hat\sigma^2 &=& \frac{-\epsilon u }{l'(y_2) - l'(y_1)} \nonumber
	\end{eqnarray}
	
	\item
	\paragraph{Poisson noise}
	In this case, we have 
	\begin{align}
		&\hat{x}_1 =  \left(y_1 + \frac{\zeta}{2}\right)\exp({\zeta l'(y_1)}) \label{poi_1}\\
		&\hat{x}_2 = \left(y_2 + \frac{\zeta}{2}\right)\exp({\zeta l'(y_2)}), \label{poi_2}
	\end{align}
	By taking the logarithm of both equations and subtracting (\ref{poi_1}) from (\ref{poi_2}), we have  
	\begin{align*}
		0 &= \log \left(1 + \frac{\epsilon u}{y_1 + \zeta/2}\right) + \zeta(l'(y_2) - l'(y_1)) \\
		&\approx \left(\frac{\epsilon u}{y_1 + \zeta/2}\right) + \zeta(l'(y_2) - l'(y_1)) ,
	\end{align*}
	where the last approximation comes from $x \approx \log(1+x)$ for sufficiently small $x$. 
	This leads to the follwoing quadratic equation for $\zeta$:
	\begin{align}
		0 &= \epsilon u + \zeta(y_1 + \zeta/2)(l'(y_2)-l'(y_1)). \label{eq:poi}
	\end{align}
	Solving quadratic equation~\eqref{eq:poi}, we can obtain the following estimate:
	\begin{eqnarray}
		\hat{\zeta} &=& \-y_1 + \sqrt{y_1^2 - 2c} \nonumber
	\end{eqnarray}
	where $c = \epsilon u/\left(l'(y_2)-l'(y_1)\right)$.
	
	\item
	\paragraph{Gamma noise}
	For the case of Gamma noise, 
	\begin{align}
		&\hat{x}_1 =  \frac{ky_1}{k- 1 -  y_1 l'(y_1)} \label{gam_1}\\
		&\hat{x}_2 = \frac{ky_2}{k- 1 -  y_2 l'(y_2)} \label{gam_2}
	\end{align}
	By taking the inverse of both equations and subtracting (\ref{gam_1}) from (\ref{gam_2}), 
	we have
	\begin{align}
		\frac{1}{y_2} - \frac{1}{y_1}& = \frac{1}{k}\left(\frac{1}{y_2} - \frac{1}{y_1} + l'(y_2)-l'(y_1)\right). \nonumber   
	\end{align}
	Then, we can obtain $\hat{k}$ by
	\begin{eqnarray}
		\hat{k} &=& 1 + \frac{l'(y_2) - l'(y_1)}{\frac{1}{y_2} - \frac{1}{y_1}}\nonumber
	\end{eqnarray}
\end{proof}

\section{Pseudocode Description}
Algorithm~\ref{algo-1} details the overall pipeline of the training procedure for the proposed method. First, the neural network $R_{\Theta}$ was trained by minimizing $\ell_{AR-DAE}(\Theta)$ to learn the estimation of the score function from the noisy input $y$. In the training, noisy images are sampled from 
an unknown noise model corrupted with various noise levels. This neural network training step is universally applied regardless of noise distribution. In particular, we annealed $\sigma_a$ from $\sigma_a^{max}$ to $\sigma_a^{min}$ to stably train the network as suggested in~\cite{song2020improved}. 
Now let $\Theta'$ be an independent copy of the parameters and after the $n^{th}$ training iteration, and we update the $\Theta'$ with the exponential moving average as indicated in Algorithm~\ref{algo-1} as suggested in~\cite{song2020improved}. 

The inference of the proposed method is described in Algorithm~\ref{algo-2}. After we obtain the trained score models $R^{\ast}_{\Theta'}$, we firstly estimate the noise model parameter $\hat{\rho}$ with Equation (10) in the main paper using Proposition 3. Once the noise model is determined, we estimate the noise level parameter for the estimated noise model   using Proposition 4. Then, the final clean image is reconstructed by Tweedie's formula as indicated in Table 2 in the main paper.   
\begin{algorithm}
	\caption{Training procedure of the proposed method}
	\label{algo-1}
	\SetKwInOut{Input}{Input}
	\SetKwInOut{kwtest}{Inference}
	\SetKwInput{Output}{Output}
	\SetKwInput{kwset}{Given}
	\kwset{learning rates  $\gamma$, number of epochs $N$;}
	\Input{noisy input $y$ from training data set $D_{\phi}$ and noise level parameter $\phi$~$\in (\sigma, \zeta, k$), neural network  $R_{\Theta}$, independent copy of the parameter $\Theta'$, annealing sigma set $S_{\sigma_a}$ with size of $T$, $S_{\sigma_a} = [\sigma_a^{min}, ..., \sigma_a^{max}]$, decay rate of exponential moving average $m$} 
	\For {$n=1$ to N}
	{
		$u \sim \mathcal{N}(0,1)$;\\
		$t \sim \mathcal{U}(0,T)$; \\
		$\sigma_a \rightarrow S_{\sigma_a}^t$ \\
		$\ell_{AR-DAE}(\Theta) ={\underset{u \sim \mathcal{N}(0,I),\sigma_a \sim \mathcal{N}(0,\delta^2)}{\underset{y \sim P_Y}{\Ed}}}\|u + \sigma_a R_\Theta(y + \sigma_a u)\|^2$;\\
		$\Theta \gets \Theta - \gamma \nabla_{\Theta} \ell_{AR-DAE}(\Theta)$;
		$\Theta' \gets m\Theta' - (1-m)\Theta$;		
	}
	\Output{Trained the score model, $ R^{\ast}_{\Theta'}(y)$ = $\hat l'(y)$ }
\end{algorithm}

\begin{algorithm}
	\caption{Inference procedure of the proposed method}
	\label{algo-2}
	\SetKwInOut{Input}{Input}
	\SetKwInOut{Output}{Output}
	\SetKwInOut{kwtest}{Inference}
	\SetKwInput{kwInit}{Noise model estiamtion}
	\SetKwInput{kwlevel}{Noise level estiamtion}
	\SetKwInput{kwset}{Given}
	\kwset{Trained score model $R_{\Theta'}^{\ast}$, the perturbed noise level $\epsilon$;}
	\Input{noisy input $y_1$ from training data set $D_{\phi}$ and noise level parameter $\phi \in (\sigma, \zeta, k)$,  and generated perturbed noisy image $y_2 = y_1 + \epsilon \mu, ~ 
		\mu \sim \mathcal{N}(0,I)$ ;}
	\kwInit{$\hat{\rho}$ by Equation (10) in the main paper}
	\kwlevel{}
	\lIf {$ 0 \leq \hat{\rho} < 0.9$ $\rightarrow$ $y \in$ \mbox{\rm Gaussian noise}}
	{$\hat{\sigma}^2 = \text{median}\left(\frac{-\epsilon u}{l'(y_2) - l'(y_1)}\right)$}
	\Output{$\hat{x} = y + \hat{\sigma}^2l'(y_1)$}
	\lElseIf {$0.9 \leq \hat{\rho} < 1.9$ $\rightarrow$ $y \in$ \mbox{\rm Poisson noise}}{$\hat{\zeta} = \text{median} \left(-y_1 + \sqrt{y_1^2 - 2c}\right)$}
	\Output{$\hat{x}$ =$\left(y+\frac{\hat{\zeta}}{2}\right)\exp^{\hat{\zeta} l'\left(y\right)}$}
	\lElseIf {$1.9 \leq \hat{\rho} < 2.9$ $\rightarrow$ $y \in$ \mbox{\rm Gamma noise} }{$\hat{k} = \text{median}\left(1 + \frac{l'(y_2) - l'(y_1)}{\frac{1}{y_2} - \frac{1}{y_1}}\right)$}	
	\Output{$\hat{x}$ = $\frac{\hat{k} y}{(\hat{k}-1)-yl'(y)}$}
	
\end{algorithm}

\section{Implementation Details}
\paragraph{Training details}
{To robustly train the proposed method, we randomly injected the perturbed noise into a noisy image instead of using linear scheduling as in \cite{kim2021noise2score}.}
In the case of the Gaussian and Gamma noise, $\sigma_a^{max}$ and $\sigma_a^{\min}$ are set to [0.1,0.001], respectively. For the Poisson noise case, $\sigma_a^{max}$ and $\sigma_a^{\min}$ was set to [0.1,0.02], respectively. 

\paragraph{Noise model estimation} 
In order to satisfy the assumption in \eqref{eq:appro}, we only calculate the pixel values that satisfies the following condition:
\begin{eqnarray}
	idx = -\epsilon <\frac{w}{\rho+b}<\epsilon \label{condition}
\end{eqnarray}
where $\epsilon$ was set to 1$\times10^{-5}$ for all of the cases. In the procedure of calculating \eqref{condition}, we can not access $\hat{\rho}$ value. Hence,
based on assumption $\rho \in (0,2)$, we empirically determine this value. In the case of additive Gaussian noise, this value is set to 2.5, otherwise to 2.2. We provide the implementation code based on Pytorch~\cite{paszke2017automatic} as shown in Listing~\ref{lst:code}.
\begin{small}
\begin{lstlisting}[caption={Source code of the proposed noise model esimation},label={lst:code}]
def noise_model_estimation(y_1,score_model):
	# Inject noise into noisy images y_1
	epsilon = 1e-5
	n = torch.randn(y_1.shape)
	noise = epsilon * n
	y_2 = y_1 + noise
	# esimate the score functions 
	l(y_1) = score_model(y_1) 
	l(y_2) = score_model(y_2) 
	# calculate each coefficient
	w = 2*(y_2*l(y_2) - y_1*l(y_1))
	a = torch.log(y_2/y_1)  
	b = (2*y_1*l(y_1))	
	# take only values under condition
	ww = w/(b+2.2)
	idx = (ww <= 1e-5) & (ww >= -1e-5)
	w = w[idx]
	b = b[idx]
	w = torch.nanmean(w)
	b = torch.nanmean(b)
	# Solve quadratic equation
	first = a*(b-2)
	second = 4*a*(- 2*a*b + w)
	sqrt = (first)**2 - second
	sqrt = torch.sqrt(sqrt)    
	p1 = (-first + sqrt)/(2*a)
	p2 = (-first - sqrt)/(2*a)
	p1 = torch.nanmean(p1)
	p2 = torch.nanmean(p2)
	# take maximum of two values
	p = max(p1,p2)
		# take maximum of p and 0
	p = max(p,0)
	return p
\end{lstlisting}
\end{small}
\paragraph{Noise level estiamtion}
To use Proposition 3, we assume that the injected small noise is sufficiently small, the equality in \eqref{mu} holds. 
To achieve this, we set $\epsilon$ to 1$\times$ $10^{-5}$ for all noise cases. 

\section{Analysis for Noise Model Estimation}
Table~\ref{tbl:ablation} shows the accuracy of the estimated noise model in the Kodak dataset. If the estimated noise distributions are equal to the truth noise distributions, we determine that the estimate is correct. We found that the proposed noise model estimation reach 100$\%$ accuracy for all cases. Thus, we concluded that we can successfully estimate the noise model with the proposed method. 

\begin{table}[ht!]
	\begin{small}
		\begin{center}
			\caption{Accuracy of the proposed noise model estimation in the Kodak dataset. }
			\vspace{-1em}
			\label{tbl:ablation}
			\resizebox{0.6\linewidth}{!}{
				\begin{tabular}{ccc}\toprule
					Noise type  & Noise level& Accuracy($\%$) \\  \cmidrule(r){1-1} \cmidrule(r){2-2} \cmidrule(r){3-3}
					\multirow{2}{*}{Gaussian}
					& $\sigma$ = 25  & 100   \\
					& $\sigma$ = 50  & 100    \\ \midrule
					\multirow{2}{*}{Poisson}
					& $\zeta$ = 0.01  & 100    \\
					& $\zeta$ = 0.05  & 100    \\ \midrule
					\multirow{2}{*}{Gamma}
					&$k$ = 100         & 100    \\
					&$k$ = 50         & 100  \\ \bottomrule
			\end{tabular}}
		\end{center}
	\end{small}
	\vspace{-2em}
\end{table}
\section{Analysis for Noise Level Estimation}
\begin{figure}[!t]
	\centering
	\begin{subfigure}{1\linewidth}
		\centering
		\includegraphics[width=0.6\linewidth]{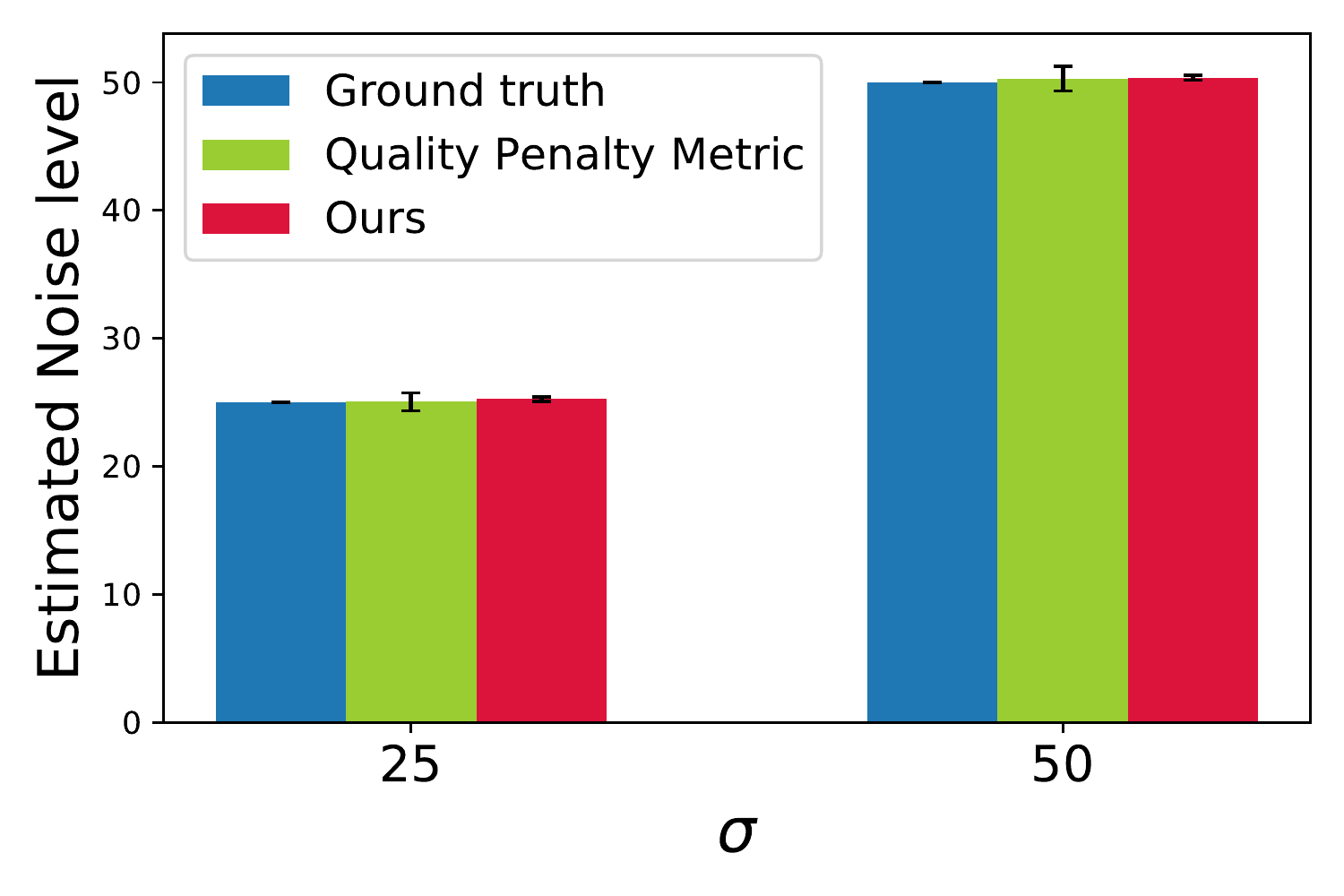} 
		\caption{} 
	\end{subfigure}
	\begin{subfigure}{1\linewidth}
		\centering
		\includegraphics[width=0.6\linewidth]{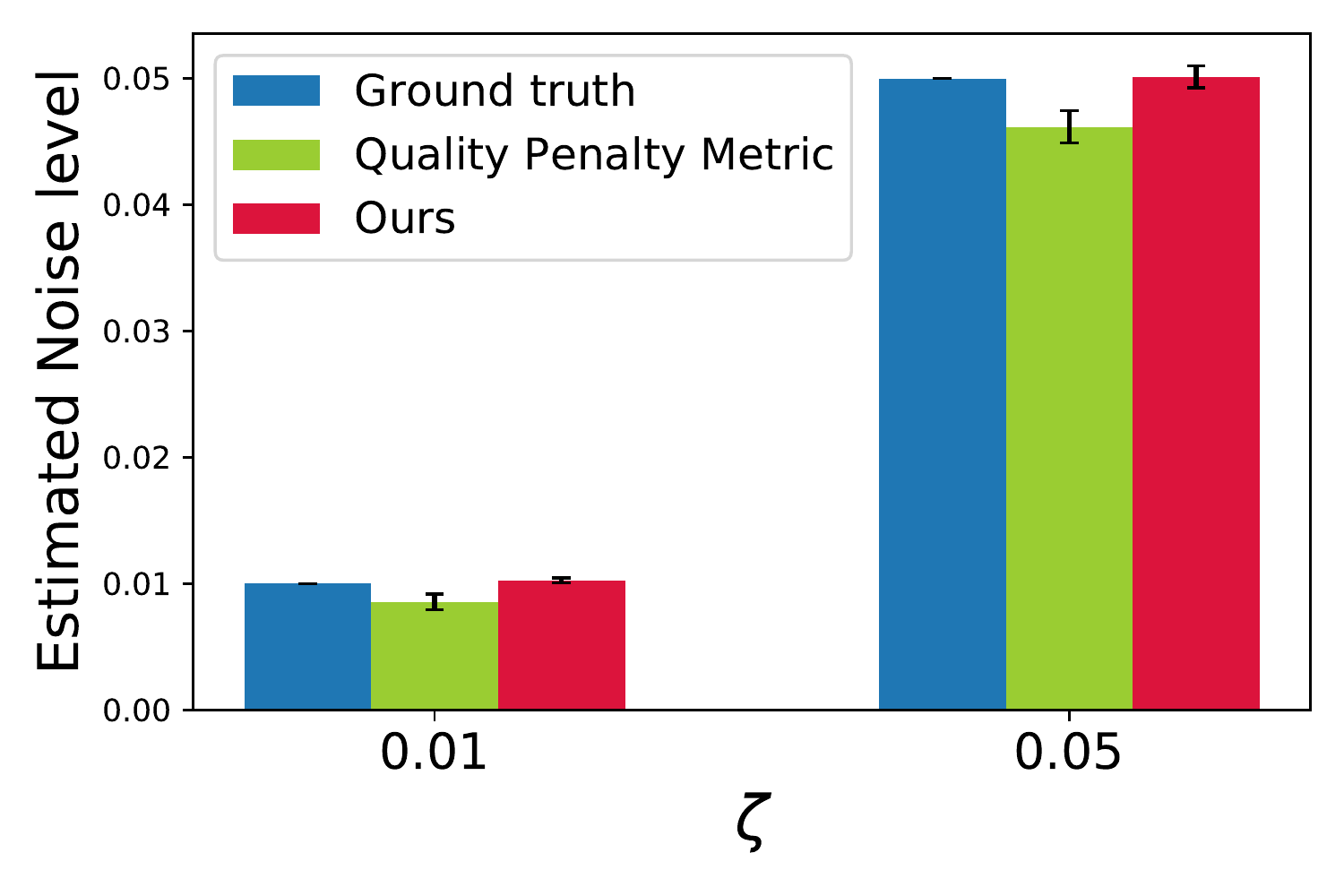}
		\caption{} 
	\end{subfigure}
	\begin{subfigure}{1\linewidth}
		\centering
		\includegraphics[width=0.6\linewidth]{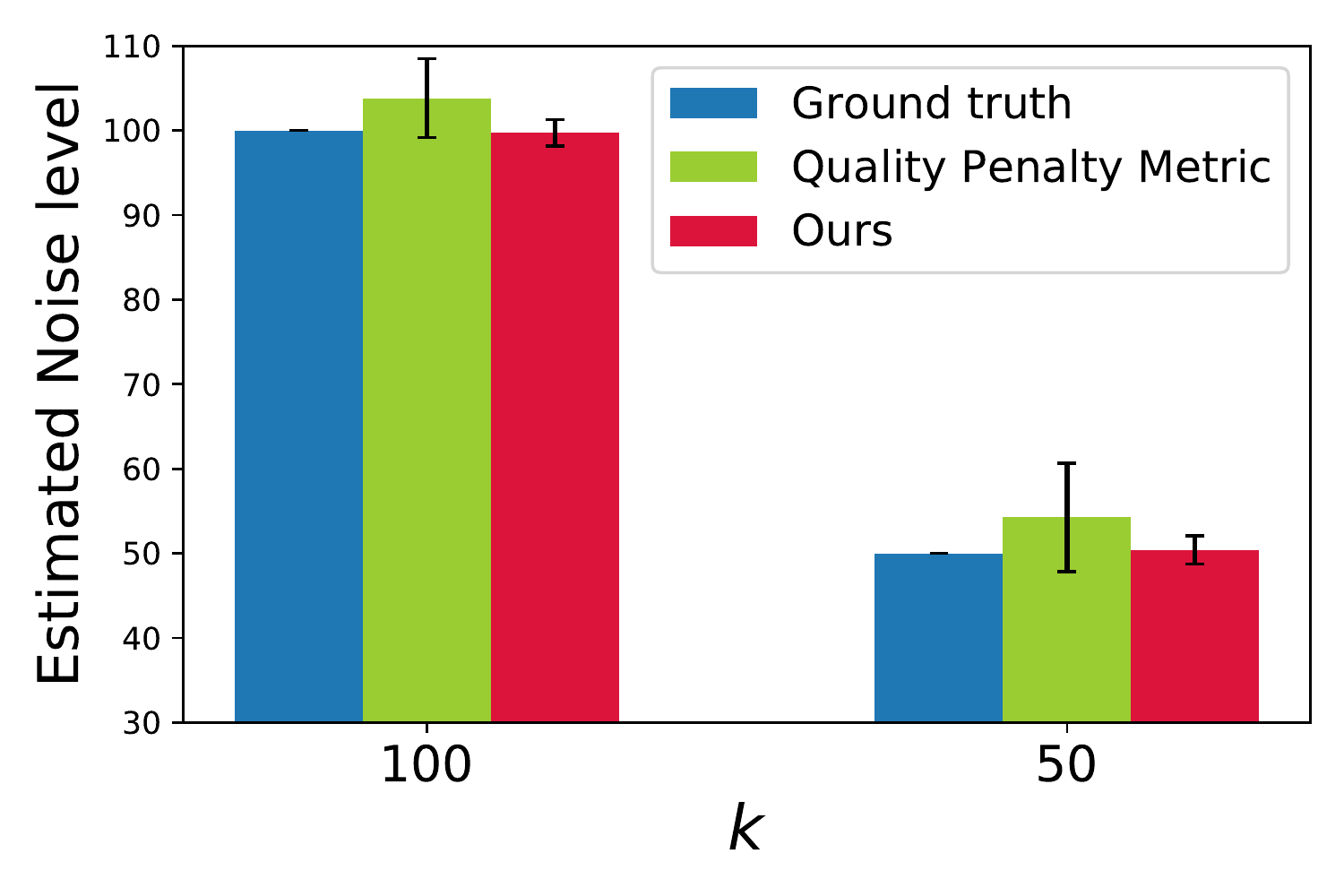}
		\caption{} 
	\end{subfigure}
	\vspace{-1em}
	\caption{Estimated noise level parameters for each noise distribution in Kodak dataset. (a) Gaussian ($\sigma$ =25, and 50). (b) Poisson ($\zeta$ = 0.01, and 0.05). (c) Gamma ($k$= 100, 50). The blue bars indicate the truth noise levels. The green bars and the red bars indicate that the average of estimated noise levels with the quality penalty metric in \cite{kim2021noise2score} and the proposed method, respectively.}
	\label{fig:noise-level}
	\vspace{-1em}
\end{figure}

Fig.~\ref{fig:noise-level} shows the bar graph of estimated noise level parameters for each noise distribution in the Kodak dataset. Similar to the ablation study in the main paper, we carried out the analysis by fixing the noise model estimation and by only varying the estimation of noise level. From the ablation study in the main paper, we expect that the quality penalty metrics  method in \cite{kim2021noise2score} estimates correctly in the case of additive Gaussian noise, but incorrectly for the case of other noise distributions. We can observe the similar findings  in Figure~\ref{fig:noise-level}. On the other hand, the proposed noise level estimation provides more accurate results compared to the quality penalty metrics in \cite{kim2021noise2score} with a small standard deviation. Thus, we can conclude that the proposed noise level estimation can successfully estimate the truth noise level in all noise distribution cases. 

\section{Ablation Study on Score Estimation}
We demonstrate the effectiveness of each component in improving the score model. Table~\ref{tbl:tech} compares the PSNR values of the results with and without each component on CBSD68 dataset (Poisson noise $\zeta$ = 0.01). EMA and GS denote Exponential Moving Average and Geometric Sequence, respectively. To fairly compare each case, we performed the ablation study using the same procedure. From the table, we observe the performance degradation when any component of the proposed method is absent. Therefore, we can conclude that EMA and GS in the proposed method are essential for improving the score model in the training procedure of the network. 
\begin{table}[h!]
	\begin{small}		
		\begin{center}
			\caption{Ablation studies on score estimation using CBSD68 data (Poisson noise cases, $\zeta = 0.01$).}
			\label{tbl:tech}
			\resizebox{0.8\linewidth}{!}{
				\begin{tabular}{ccccc}\toprule
					Component  & Ours & Case1  & Case2 & Case3 \\ \cmidrule(r){1-1} \cmidrule(r){2-2} \cmidrule(r){3-5} 
					EMA   & $\cmark$  & $\cmark$  & $\xmark$ & $\xmark$ \\\cmidrule(r){1-1} \cmidrule(r){2-2} \cmidrule(r){3-5} 
					GS    & $\cmark$ & $\xmark$ & $\cmark$ & $\xmark$ \\ \cmidrule(r){1-1} \cmidrule(r){2-2} \cmidrule(r){3-5} 
					PSNR(dB) & \textbf{32.53} & 32.41  & 32.23 & 32.03 \\ \bottomrule
			\end{tabular}}
		\end{center}
	\end{small}
	\vspace{-1em}
\end{table}
\begin{figure*}[!b]
	\centering
	\includegraphics[width=1\linewidth]{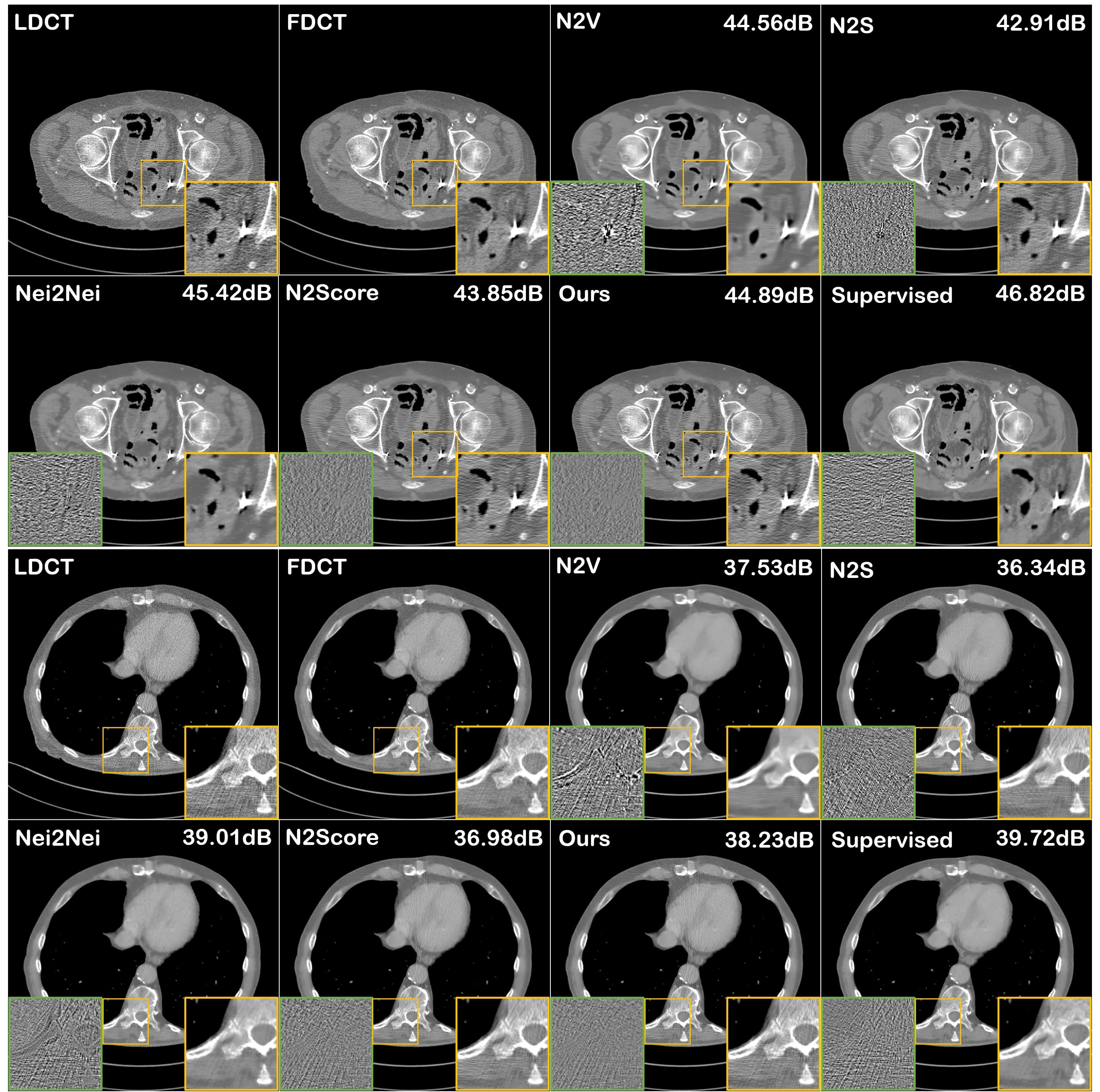}  
	\caption{Denoising results of AAPM data using various methods. The yellow box and green box show the enlarged view of image and difference image between network input and output, respectively. The intensity window of CT image is (-500,500)[HU] and the intensity window of difference is (-200,200) [HU].}
	\label{fig:ct}
\end{figure*}
\section{Qualitative Results }
We provide more examples of the denoising results by various methods on the AAPM CT dataset~\cite{mccollough2017low} as shown in Fig.~\ref{fig:ct}. Similar to the results shown in the main paper, the improvements are consistent. In particular,  other self-supervised approaches produce over-smooth denoised images,
which produces the boundary structure of CT images in difference images. On the other hand, the proposed method provides similar results compared to target images and only generates noise components in the difference images.

\section{Limitation and Negative Societal Impacts }
While we provide a unified approach for noise distribution adaptive self-supervised image denoising, the method is not free of limitations. 
As the formulae for estimating of noise model and levels were derived from several approximations, it could fail in a real-world dataset. Furthermore, the noise model may not be described by Tweedie distribution in real environments. 

As a negative societal impact, the failure of image denoising methods can lead to side effects. For example, removing both the noise and the texture of the medical images could lead to misdiagnosis.

\end{document}

%% file: macros.tex
\long\def\comment#1{} 

\newcommand{\Ed}{{{\mathbb E}}}

\newcommand{\beq}{\begin{equation}}
\newcommand{\eeq}{\end{equation}}
\newcommand{\beqa}{\begin{eqnarray}}
\newcommand{\eeqa}{\end{eqnarray}}

